\documentclass{llncs}
\usepackage[bookmarks=true,hyperfigures=true,colorlinks=true,linkcolor=blue,citecolor=blue,backref=page,pagebackref=false,pdfpagemode=fullscreen,plainpages=false,pdfpagelabels]{hyperref}
\usepackage[all]{hypcap}
\usepackage{mathptmx}      
\usepackage[sort]{natbib}
\bibpunct{(}{)}{;}{a}{}{,}
\usepackage{txfonts}
\usepackage{caption,url,latexsym,alltt}
\usepackage[utf8]{inputenc}
\DeclareUnicodeCharacter{25A1}{\ensuremath\Box}
\DeclareUnicodeCharacter{25C7}{\ensuremath\Diamond}
\newcommand{\wig}{\(\sim\)}
\newcommand{\spc}{\(\,\)}
\newcommand{\defdef}{\mbox{$\stackrel{\rm def}{=}$}}
\newcommand{\defn}{\mathrel{\defdef}}
\newcommand{\hozline}{{\noindent\rule{\textwidth}{0.2mm}}}
\newlength{\hsbw}
\def\sessionsize{\small}
\newenvironment{session}{\begin{flushleft}
 \setlength{\hsbw}{\linewidth}
 \addtolength{\hsbw}{-\arrayrulewidth}
 \addtolength{\hsbw}{-\tabcolsep}
 \begin{tabular}{@{}|c@{}|@{}}\hline 
 \begin{minipage}[b]{\hsbw}
 \begingroup\sessionsize\vspace*{1.2ex}\begin{alltt}}{\end{alltt}\endgroup\end{minipage}\\[1ex] \hline
 \end{tabular}
 \end{flushleft}}
\newenvironment{sessionlab}[1]{\begin{flushleft}
 \setlength{\hsbw}{\linewidth}
 \addtolength{\hsbw}{-\arrayrulewidth}
 \addtolength{\hsbw}{-\tabcolsep}
 \begin{tabular}{@{}|c@{}|@{}}\hline 
 \begin{minipage}[b]{\hsbw}
 \vspace*{-.5pt}
 \begin{flushright}
 \rule{0.01in}{.15in}\rule{0.9in}{0.01in}\hspace{-0.95in}
 \raisebox{0.04in}{\makebox[0.9in][c]{\footnotesize #1}}
 \end{flushright}
 \vspace*{-.47in}
 \begingroup\small\vspace*{2.0ex}\begin{alltt}}{\end{alltt}\endgroup\end{minipage}\\ \hline 
 \end{tabular}
 \end{flushleft}}
\newenvironment{sessionbiglab}[1]{\begin{flushleft}
 \setlength{\hsbw}{\linewidth}
 \addtolength{\hsbw}{-\arrayrulewidth}
 \addtolength{\hsbw}{-\tabcolsep}
 \begin{tabular}{@{}|c@{}|@{}}\hline 
 \begin{minipage}[b]{\hsbw}
 \vspace*{-.5pt}
 \begin{flushright}
 \rule{0.01in}{.15in}\rule{1.45in}{0.01in}\hspace{-1.5in}
 \raisebox{0.04in}{\makebox[1.45in][c]{\footnotesize #1}}
 \end{flushright}
 \vspace*{-.47in}
 \begingroup\small\vspace*{2.0ex}\begin{alltt}}{\end{alltt}\endgroup\end{minipage}\\ \hline 
 \end{tabular}
 \end{flushleft}}
\begin{document}
\author{John Rushby}
\title{Mechanized Analysis of Anselm's\\ Modal Ontological Argument}
\institute{
  Computer Science Laboratory, SRI International, Menlo Park CA 94025 USA\\
              \email{Rushby@csl.sri.com}}
\maketitle
\pagestyle{plain}
\begin{abstract}
We use a mechanized verification system, PVS, to examine the argument
from Anselm's Proslogion Chapter III, the so-called ``Modal Ontological
Argument.''  We consider several published formalizations for the
argument and show they are all essentially similar.  Furthermore, we
show that the argument is trivial once the modal axioms are taken into
account.

This work is an illustration of computational philosophy and, in
addition, shows how these methods can help detect and rectify errors
in modal reasoning.

\end{abstract}

\setcounter{section}{-1}
\section{Preamble}

This is a minor update with better typesetting and some small addenda
plus a new (June 2024) postscript
to a paper published in the International Journal for Philosophy of
Religion, vol. 89, pp. 135--152, April 2021 (online publication 4
August 2020).

\section{Introduction}

The Ontological Argument is a proof of the existence of God presented
by St.\ Anselm of Canterbury in his \emph{Proslogion} 
\citep{Anselm:Proslogion}.  It has been studied and debated by
philosophers and theologians ever since.

The Ontological Argument has traditionally been identified with
Chapter II of the Proslogion, but in the middle years of the last
century philosophers noted that Chapter III presents an argument that
could be interpreted as an alternative proof for God's existence.
This discovery is generally attributed to \cite{Malcolm60} although
Hartshorne had made similar observations earlier.  (\citet[Part 1,
Section 8]{Hartshorne:discovery} gives some of the history, and also
\citet[Introduction]{Smith:Anselm}.)  The discovery was that the
argument of Chapter III can be seen as an independent, self-contained
proof; previously, and for many commentators still today
\citep{Campbell18:book, Campbell19:AnselmP3} \citep[p.\ 82]{Sobel03},
Chapter III is seen as a continuation of the argument of Chapter II.

We reproduce an English translation of Chapter III in Figure
\ref{text}.  Its argument, which is found in its second paragraph and
highlighted in italics, is couched in terms of beings whose existence
is \emph{necessary} (i.e., ``that which cannot be conceived not to
exist'') and has come to be known as Anselm's \emph{Modal} Argument,
since qualifiers such as ``necessary'' (with dual ``possible'')
indicate \emph{modes} of truth.  Philosophers since Aristotle have
attempted to construct specialized logics for reasoning about these
but modern \emph{modal logics} were developed only in the 20th century
\citep{Cresswell&Hughes96}.

\begin{figure}[tp]
\begin{verse}
God cannot be conceived not to exist. God is that,
than which nothing greater can be conceived. That which can be
conceived not to exist is not God.

And it assuredly exists so truly, that it cannot be conceived not to
exist.  \emph{For, it is possible to conceive of a being which cannot be
conceived not to exist; and this is greater than one which can be
conceived not to exist. Hence, if that, than which nothing greater can
be conceived, can be conceived not to exist, it is not that, than
which nothing greater can be conceived. But this is an irreconcilable
contradiction. There is, then, so truly a being than which nothing
greater can be conceived to exist, that it cannot even be conceived
not to exist; and this being you are, O Lord, our God.}

So truly, therefore, do you exist, O Lord, my God, that you can not be
conceived not to exist; and rightly. For, if a mind could conceive of
a being better than you, the creature would rise above the Creator;
and this is most absurd. And, indeed, whatever else there is, except
you alone, can be conceived not to exist. To you alone, therefore, it
belongs to exist more truly than all other beings, and hence in a
higher degree than all others. For, whatever else exists does not
exist so truly, and hence in a less degree it belongs to it to
exist. Why, then, has the fool said in his heart, there is no God
(Psalms xiv. 1), since it is so evident, to a rational mind, that you
do exist in the highest degree of all? Why, except that he is dull and
a fool?
\end{verse}
\vspace*{-1ex}
\hozline
\caption{\label{text}The Complete Chapter III of St.\ Anselm's Proslogion \citeyearpar{Anselm:Proslogion}.}
\vspace*{-3ex}
\end{figure}

\vspace*{-2ex}
\subsubsection*{Addendum}$\!$to the published paper (8 June 2023).

The published paper provides no description of the path from Anselm's
informal presentation of the argument in Figure \ref{text} to the
formal versions considered later.

The first step in doing this is to introduce $g$ as an abbreviation for (the
existence of) ``that than which nothing greater can be conceived'' and
to use $\Diamond$ to mean ``it is conceivable that.''  Then the
sentence ``Hence, if that, than which nothing greater can be
conceived, can be conceived not to exist, it is not that, than which
nothing greater can be conceived'' can be read as $\Diamond \neg g
\supset \neg g$, and the contrapositive of this is $g \supset \neg
\Diamond \neg g$ (see first footnote for notation).

At this point we need to decide how to interpret $\Diamond$ formally
and our choice is to treat it as the ``possibly'' qualifier of Alethic
modal logic.  Then $\neg \Diamond \neg$ becomes the ``necessary''
qualifier $\Box$ and the formal rendition becomes $$g \supset \Box
g.$$

This has become known as ``Anselm's Principle'' and, in modern
language, its justification might be that if a greatest being exists,
then its existence could not be contingent (i.e., chance), because we
can then conceive of a greater being (i.e., one whose existence is
necessary).  Hence if a greatest being exists, then its existence must
be necessary.

Next, it seems implicit that the existence of $g$ is conceivable,
and this is rendered as $$\Diamond g.$$

We now have the two premises of the argument, and its completion
requires rules for reasoning about the modal qualifiers $\Diamond$ and
$\Box$.  When sound treatments for modal logics started to be
developed (in the last century), it became apparent that additional
axioms were needed to describe properties of the modal qualifiers, and
that different axioms were required according to the different
interpretations of the qualifiers (e.g., knowledge vs.\ belief).
Formalizations of the
ontological argument use an Alethic logic (one in which the
qualifiers are interpreted as ``possibly'' and ``necessary'' or, to
be more faithful to Anselm ``it is conceivable that'' and ``it is
inconceivable that not'') and suitable auxiliary axioms are those
known as T, B, and 5, which are described in Section \ref{pml}.\\
\textbf{End of Addendum}\\[-1ex]

\sloppypar

In this paper, I subject several published formalized variants of
Anselm's Modal Argument to mechanized analysis in the PVS Verification
system; specifically, I consider versions by \citet[Section
4.1]{Eder&Ramharter15}, \cite{Kane84}, \citet{Malcolm60} as reported
by \citet{Hartshorne61}, \citet{Adams71} as reported by
\citet{Matthews:ontological}, and \citet[pp.\
50--51]{Hartshorne:perfection}.  This work is an elementary example of
\emph{Computational Philosophy} \citep{SEP-Computational-Philosophy},
that is, application of mechanized computation and reasoning to
philosophical topics.  

There are several other applications of
automated reasoning to ontological arguments, including first and
higher-order treatments of Anselm's traditional argument
\citep{Rushby19:ontargbegsvac,Oppenheimer&Zalta11,Rushby:ontological13},
modal treatments of that argument \citep{Rushby:modalont19}, and
G\"{o}del's modal argument \citep{Benzmueller&Paleo:ECAI14}.  These
other applications use moderately advanced logical constructions, such as
definite descriptions
\citep{Oppenheimer&Zalta11,Rushby:ontological13}, higher order logic
\citep{Rushby19:ontargbegsvac}, and first order modal logic
\citep{Rushby:modalont19,Benzmueller&Paleo:ECAI14}, whose mechanized
support is fairly intricate and prone to errors \citep{Garbacz12}.

In contrast, Anselm's modal argument requires only propositional modal
logic.  However, this is mechanized by embedding it in the classical
logic of a verification system, which requires some care.
Nonetheless, these are among the simplest examples of Computational
Philosophy and can serve as an introduction to the field.  I will
show, however, that despite its simplicity, this exercise delivers
useful and novel results of both a positive and negative kind---the
latter being to expose a lamentable error in a paper published in this
very journal.

The structure of the paper is as follows.  Section 2 provides a short
introduction to the embedding of propositional modal logic in PVS,
which is the system that provides the mechanization used here; the
five variants of the Argument are presented and analyzed in Section 3;
Section 4 discusses the value of this exercise and some related
analyses, and Section 5 provides conclusions.

\section{Propositional Modal Logic in PVS}
\label{pml}

Modal logics allow reasoning about various \emph{modes} of truth: for
example, what it means for something to be \emph{necessarily} true, or to
\emph{know} that something is true as opposed to merely
\emph{believing} it.  

The modal \emph{qualifier} $\Box$ and its dual $\Diamond$ (defined as
$\neg \Box \neg$)\footnote{We use $\neg$ for negation, $\wedge$ for
conjunction, $\vee$ for disjunction and $\supset$ for material implication.} are used to
indicate expressions that should be interpreted modally.  All modal
logics share the same basic structure but they employ different sets
of axioms and make other adjustments according to the mode attributed
to the qualifiers.  For example, in an \emph{Alethic} modal logic,
where $\Box$ is interpreted as necessity, we will expect the formula
$\Box P \supset P$ to hold: if something is necessarily true, then it
should be true.  But we would not expect this formula to hold in a
\emph{Doxastic} logic, where $\Box$ is interpreted as belief.
Instead, we might expect $\Box P \supset \Diamond P$ to hold: if I
believe P is true, then I cannot also believe it to be false
(reading $\Diamond P$ as $\neg \Box \neg P$).  There is a collection
of formulas such as these that have standard names (the two above are
called T and D, respectively) and that are used in various
combinations to axiomatize different modal logics.  Some of the common
ones are listed below.
\begin{description}
\item[K:] $\Box (P \supset Q) \supset (\Box P \supset \Box Q)$ [this formula
    is true in all modal logics],
\item[T:] $\Box P \supset P$,
\item[D:] $\Box P \supset \Diamond P$,
\item[B:] $P \supset \Box\Diamond P$,
\item[4:] $\Box P \supset \Box \Box P$,
\item[5:] $\Diamond P \supset \Box\Diamond P$.
\end{description}

The semantics of modal logics are interpreted relative to a set of
\emph{possible worlds}, so that $\Box P$ means true in \emph{all}
worlds and $\Diamond P$ means true in \emph{some} world.  To represent
different interpretations for the modalities, we impose an
\emph{accessibility} relation on possible worlds and refine the
statements above to true in \emph{all accessible} and \emph{some
accessible} worlds, respectively.

There is a relationship between the standard modal axioms and
properties of the accessibility relation.  For example, T and D,
mentioned above, correspond to accessibility relations that are
reflexive and serial, respectively.  The modal ontological argument is
expressed in Alethic modal logic, and such logics have accessibility
relations that are equivalence relations.  This corresponds to the
combination of modal axioms T (reflexive) and either 4 (transitive)
plus B (symmetric), or 5 (Euclidean).\footnote{The modal logic with no
accessibility relation has the same theorems as one whose
accessibility relation is an equivalence.}

We will undertake our examination of the modal ontological argument
using the PVS Verification System.  Verification systems are tools
from computer science that are generally used for exploration and
verification of software or hardware designs and algorithms; they
comprise a specification language, which is essentially a rich logic,
and a collection of powerful deductive engines (e.g., satisfiability
solvers for combinations of theories, model checkers, and automated
and interactive theorem provers).  In particular, PVS has a
specification language based on higher-order logic and its proof
automation is guided interactively.  It is generally applied to
analysis and verification of computational systems and has more than
4,000 citations.  We do not describe PVS in detail here but we do try
to provide enough information to make this presentation self
contained; see \cite{PVS} for the PVS system description and
\cite{Rushby:ontological13,Rushby18:begging,Rushby19:ontargbegsvac,Rushby:modalont19}
for previous applications to the (traditional, Proslogion Chapter II)
Ontological Argument.

\begin{figure}[t]
\begin{sessionlab}{PVS}
shallow_pml: THEORY
BEGIN

  worlds: TYPE+
  access: pred [[worlds, worlds]]
  pmlformulas: TYPE = [worlds -> bool]
  pvars: TYPE+

  v, w: VAR worlds
  x, y: VAR pvars
  P, Q: VAR pmlformulas

  val(x)(w): bool

  \wig(P)(w): bool = NOT P(w) ;
  &(P, Q)(w): bool = P(w) AND Q(w) ;
  \verb|\|/(P, Q)(w): bool = P(w) OR Q(w) ;
  =>(P, Q)(w): bool = P(w) IMPLIES Q(w) ;

  □(P)(w): bool = FORALL v: access(w, v) IMPLIES P(v) ;
  <>(P)(w): bool = EXISTS v: access(w, v) AND P(v) ;
  dexpand: LEMMA <>\,P = \wig\,□\,\wig\,P ;

  |=(w, P): bool = P(w)
  valid(P): bool = FORALL w: w |= P

  validval(x: pvars): bool = valid(val(x))
  CONVERSION valid, val

END shallow_pml
\end{sessionlab}
\caption{\label{embed}Shallow Embedding of Propositional Modal Logic
in PVS.}
\end{figure}

There is a standard translation from modal logic to classical first-
or higher-order logic \citep{standard-translation} and we use this to
provide a \emph{shallow embedding} into the dependently-typed
higher-order classical logic of PVS in a way that preserves much of
its proof automation.  The embedding is described in tutorial detail
elsewhere \citep{Rushby:Modal17}.  The core of the embedding is shown
in Figure \ref{embed}; the basic idea is that modal expressions are
``lifted'' so that they become functions on worlds.  The function
\texttt{val(x)(w)} provides the valuation for propositional constant
\texttt{x} in world \texttt{w}, and \texttt{\wig}, \texttt{\&},
\verb|\/|, and \texttt{=>} provide the ``lifted'' versions of
negation, conjunction, disjunction, and material implication,
respectively.  We then define the modalities as described above, and
define modal validity as truth in all worlds.

When we write a modal formula such as \texttt{<>\,P => P} in PVS, we
really intend its classical embedding, \texttt{valid(<>\,P => P)}, and
this is taken care of automatically using the PVS \texttt{CONVERSION}s
(see \cite{Rushby:Modal17}) specified at the bottom of the theory, and
in some of the theories that appear later.

\vfill

\section{Reconstructions of the Modal Ontological Argument}
\label{modal-arg}

We will examine five reconstructions of Anselm's modal argument.  As
we will discover, they are all variants of the same rather trivial
deduction.

\subsection{The Version by Eder and Ramharter}
\label{er}

I will begin with the version of \citet[Section
4.1]{Eder&Ramharter15}.  They attribute this to \citet[pp.\
49--51]{Hartshorne:perfection} but my reading of Hartshorne's version
is somewhat different and I will examine that
later.\footnote{Hartshorne uses the arrow $\rightarrow$ to mean
\emph{strict} implication, so his version of \cite[pp.\
49--51]{Hartshorne:perfection} is that examined in Section
\ref{hart1}.  In the published paper, I stated ``elsewhere, \citet[p.\
97]{Hartshorne:discovery} does state `Anselm's Principle' in the form
used by Eder and Ramharter.''  This is incorrect, however, and I am
grateful to Diego Murcia for pointing out that Hartshorne uses strict
implication in that version, too.}

The argument is presented in an \emph{Alethic} modal
logic: that is, one where the modalities express \emph{necessity}
($\Box$) and \emph{possibility} ($\Diamond$), respectively.

\begin{description}\label{erarg}
\item[Definition:] $g$ is an abbreviation for  ``there exists a being than
 which there is none greater'' (i.e., God).

\item[Premise ER1:] $g \supset \Box g$ (i.e., if God exists, then he
exists necessarily).  \par \noindent Eder and Ramharter refer to ER1
as \emph{Anselm's Principle} and state that a more precise reading
(i.e., a better match to Anselm's language) is ``if God exists, then
his nonexistence is inconceivable'' (i.e., reading $\Box g$ as $\neg
\Diamond \neg g$).

\item[Premise ER2:] $\Diamond g$ (i.e., God's existence is possible).

\item[Conclusion ERC:] $g$ (i.e., God exists in a classical sense).

\end{description}

Before we proceed to examine the logic of this formulation, we note
that it is abstracted quite significantly from Anselm's natural
language description as given (in English translation) in the second
paragraph of Figure \ref{text}.  Anselm speaks of beings that can be
``conceived'' to exist, and also has a relation of ``greater'' among
beings.  Eder and Ramharter's paper is largely concerned with what it
means for a logical formulation to be a good reconstruction of an
informal argument and they note that this formulation fails several of
their criteria.  We will return to some of these points in Section
\ref{discuss}.

\begin{figure}[t]
\begin{sessionlab}{PVS}
Eder_Ramharter: THEORY
BEGIN

IMPORTING simple_pml

  g: pvars
  P: VAR pmlformulas

  ER1: AXIOM g => □\,g

  ER2: AXIOM <>\,g

  ER_triv: LEMMA symmetric?(access)\,\,IMPLIES (P => □\,P)\,\,IMPLIES (<>\,P => P)

  CONVERSION+ validval

  ERC_tmp: THEOREM g

  ERC: THEOREM symmetric?(access) => g

END Eder_Ramharter
\end{sessionlab}
\caption{\label{er-pvs}Eder and Ramharter's Version in PVS.}
\end{figure}

Eder and Ramharter's formulas ER1, ER2, and ERC are all rendered
straightforwardly into PVS as shown in Figure \ref{er-pvs}.  The
imported theory \texttt{simple\_pml} provides the shallow embedding of
propositional modal logic shown in the previous section, plus some
additional theories including the standard modal axioms.

\subsubsection{Automated Proof.}
\label{automated}

The formula \texttt{ERC\_tmp} states the basic conclusion, which we
attempt to prove using the following PVS proof commands.
\begin{sessionbiglab}{PVS proof steps}
(lemma "ER1") (lemma "ER2")
(grind :if-match nil)
(inst?)
(grind :polarity? t)
\end{sessionbiglab}
These install the two premises, then invoke the standard automated
proof command \texttt{grind} with instantiation of variables disabled
(this just unwinds the embedding of modal logic); next we perform
heuristic instantiation on one of the premises, and finally use
\texttt{grind} again, with variable instantiation instructed to pay
attention to the polarity of variables.

This delivers the following
proof sequent.
\begin{sessionlab}{PVS sequent}
\{-1\}  access(w!1, v!1)
\{-2\}  val(g)(v!1)
  |-------
\spc\{1\}   access(v!1, w!1)
[2] \(\!\)  val(g)(w!1)
\end{sessionlab}

The interpretation of a PVS sequent is that the conjunction of
formulas above the \texttt{|-------} turnstile line should entail the
disjunction of those below.  Terms such as \texttt{w!1} are Skolem
constants, \texttt{access} is the accessibility relation on possible
worlds, and \texttt{val} is the valuation function for propositional
constants, so \texttt{val(g)(v!1)} is the valuation of \texttt{g} in
world \texttt{v!1}.  Here, we see that the proof would be completed if
line -1 implied line 1: that is, if the accessibility relation were
symmetric.

Thus, we recognize that the conclusion to the argument should be
modified to mention symmetry, and this explains the final form for
\texttt{ERC} shown in Figure \ref{er-pvs}, which is proved by the same
proof steps as above.

Alternatively, we could cite the standard modal axiom B (i.e., $P
\supset \Box \Diamond P$), which characterizes symmetric accessibility
relations, as an additional premise.  In this case, \texttt{ERC\_tmp}
is proved by the following commands.  Notice that the instantiation
into B is not trivial (observe the negation).
\begin{sessionlab}{PVS proof steps}
(lemma "ER1") (lemma "ER2")
(lemma "B") (inst -1 "\wig{}val(g)")
(grind :if-match nil)
(inst? -2)
(grind :polarity? t)
\end{sessionlab}

As can be inferred from the PVS sequent shown on a previous page,
these proofs operate directly on the possible world semantics of modal
logic: essentially they expand the definitions given in Figure
\ref{embed} and then perform quantificational reasoning over the
possible worlds.  This is automated effectively and efficiently by the
standard proof mechanization of PVS and the intermediate proof states,
displayed as sequents, are fairly easy to interpret.\footnote{Greater
automation is possible using modern SMT solvers
\citep{Ge&DeMoura09:quantSMT} but that can be less helpful when, as
here, a proof fails and we want to understand why.} However, it does
not reproduce the style of semi-formal proofs that typically accompany
journal presentations.  Accordingly, we now illustrate how PVS can be
used to examine those kinds of proofs.

\vspace*{-2ex}
\subsubsection*{Addendum}$\!$to the published paper (3 June 2024).

It is worth noting the relationship between Anselm's Principle as
expressed in the premise ER1 and the rule of \emph{necessitation} in
modal logic, generally named N:
\begin{description}
\item[N:] if $P$ is valid, so is $\Box P$.
\end{description}
This is actually a metatheorem, true in all modal logics; it
says that if $P$ is just plain true (i.e., without reference to a
possible world), then it is true in all possible worlds.

Depending on the syntactic conventions available, N can be
written to look very similar to ER1.  For example, in PVS,
N (instantiated for g) can be written
\begin{sessionlab}{N in PVS}
  N: LEMMA g IMPLIES □ g
\end{sessionlab}
However, if we expand the \texttt{CONVERSION}s applied in this lemma (using the
PVS function \texttt{prettyprint-expanded}), we see that it is
\begin{sessionbiglab}{N expanded in PVS}
  N: LEMMA valid(g) IMPLIES valid(□ g)
\end{sessionbiglab}
whereas doing the same for ER1 delivers the following.
\begin{sessionbiglab}{ER1 expanded in PVS}
  ER1: AXIOM valid(g => □ g)
\end{sessionbiglab}

The crucial difference in modal scope (and hence quantification) is
now quite evident.  N says that if g is true in every world, then it
must be true in every world that is accessible from a given world.
This is obviously true (and provable).  Whereas ER1 says that if P is
true in a given world, it must also be true in every world that is
accessible from that world.  This is a strong claim, plausible for
God, but not in general.

This topic highlights another one worth noting, namely, that the deduction
theorem is not valid in modal logic.  That is to say, the following
(which would allow \texttt{ER1} to be derived from \texttt{N}) is not
provable (its converse is, however).\label{nonded}
\begin{sessionlab}{PVS fragment}

  nondeduction: CLAIM (valid(P) IMPLIES valid(Q)) IMPLIES valid(P => Q)
\end{sessionlab}

Related to this is the observation that it is often necessary to be
very careful about which parts of a sentence are to be interpreted
modally, and which are conventional propositional logic.  In Section
\ref{discuss} we will see errors due to this issue.\\
\textbf{End of Addendum}\\[-1ex]

\subsubsection{Reconstructing a Manual Proof.}
\label{manual}

\citet[Section 4.2]{Eder&Ramharter15} provide a typical semi-formal
manual proof that is presented below in a slightly reorganized form.
\begin{description}
\item[Step 1:] Applying Modal Axiom 5 to $\neg g$ and interpreting
$\Diamond$ as $\neg \Box \neg$ we obtain\\ $\neg \Box g \supset \Box
\neg \Box g$; standard propositional logic then gives $\Box g \vee
\Box \neg \Box g$.

\item[Step 2:] By contraposing Anselm's principle ER1
(i.e., $\neg □ g \supset \neg g$)
and instantiating Modal Axiom K with $P = \neg \Box g$ and $Q = \neg g$
we arrive at $ \Box \neg \Box g \supset \Box \neg g$.

\item[Step 3:] The previous steps combine to give $\Box g \vee \Box
\neg g$
which can be rewritten as\\ $\Box g \vee \neg \Diamond \neg \neg g$
and then simplified to $\Box g \vee \neg \Diamond g$.

\item[Conclusion:] The previous step combined with God's possibility ER2
allows us to conclude $\Box g$ from which Modal Axiom T allows us to
infer $g$.
\end{description}

We can reproduce this level of reasoning in PVS by stating the
conclusion of each step as a lemma as shown below.

\begin{sessionlab}{PVS text}
step1: LEMMA □\,g \verb|\|/ □\,\wig{}□\,g

step2: LEMMA □\,\wig{}□\,g => □\,\wig{}g

step3: LEMMA □\,g \verb|\|/ □\,\wig{}g
\end{sessionlab}

We then invite PVS to prove each of the lemmas and the conclusion
\texttt{ERC\_tmp} without expanding the modal qualifiers.  Here is the
beginning of the proof for \texttt{step1}.
\begin{sessionlab}{PVS proof steps}
 (lemma "five") (inst - "\wig{}g")
 (grind :if-match nil :exclude ("<>" "□") :rewrites "dexpand")
 (inst?)
\end{sessionlab}

The \texttt{:exclude} construct prevents expansion of the modal
qualifiers, but note that \texttt{dexpand} will automatically rewrite
$\Diamond$ as $\neg \Box \neg$.  This brings us to the following
sequent, which represents the instantiation of Modal Axiom 5 (the
formula above the turnstile line) and allows us to see how this will
discharge the conclusion to \texttt{step1} (the two formulas below the
line).  That final step is accomplished by the proof command
\texttt{(grind)}.  Note that the Skolem constant on worlds
\texttt{w!1} comes from the definition of validity for modal
constructs as true in all worlds.
\begin{sessionlab}{PVS sequent}
\{-1\}  NOT (□\wig\wig{}val(g))(w!1) IMPLIES (□\wig□\wig\wig{}val(g))(w!1)
  |-------
\{1\}   (□val(g))(w!1)
\{2\}   (□\wig□val(g))(w!1)
\end{sessionlab}

Preference for one or the other style of mechanically assisted proof
depends on the purpose of the exercise.  If it is to examine validity
and soundness of a particular argument (i.e., selection of premises),
then I think the automated approach is preferable: it is generally
fast and simple.  But if the purpose is to examine a specific proof
(i.e., chain of inferences) then a mechanically assisted
reconstruction will be the necessary choice.  I will examine some
specific proofs in Section \ref{discuss} but will favor automated
verification for the remainder of this section.  

Returning to the automated verification of Section \ref{automated},
PVS assures us that this specification of the argument is valid, so we
should now consider whether the premises and assumptions are
reasonable and sound.

\subsubsection{Analysis and Interpretation.}
\label{analysis}

We focus first on the assumption of a symmetric accessibility relation
or, alternatively, the standard axiom B.

The argument makes sense only in an Alethic logic, that is one where
$\Box$ and $\Diamond$ are interpreted as \emph{necessary} and
\emph{possibly}, respectively, and Alethic modal logics are
characterized by having accessibility relations that are equivalence
relations.  Hence, the assumption of symmetry in \texttt{ERC} presents
no difficulty.  Dually, Alethic logics are also characterized by the
standard axioms T, 4, and 5---a combination that is generally referred
to as S5\footnote{The modal formula K is often mentioned, too, and the
logic is then sometimes referred to as KT45; however, K is a theorem,
true in all modal logics (alternatively, it can be seen to
characterize what is a modal logic), so its mention is redundant;
furthermore, T and 5 imply 4, so mention of 4 is also
redundant.}---from which B can be derived as a theorem.  We conclude
that the assumption of symmetry or modal axiom B is uncontroversial,
provided one accepts that the Alethic logic needed here is correctly
characterized by S5---which few would dispute.  If in doubt,
\citet{Ladstaetter12} provides a clear description of \emph{logical},
\emph{metaphysical}, and \emph{physical} notions of necessity, and
explains why logical necessity is required here, and why S5 is the
right formalization.

We note that the manual proof of Section \ref{manual} used Modal
Axioms 5, K, and T instead of B.  However, K is true of all modal
logics, and 5 and T are included in S5, and therefore this combination
is acceptable in an Alethic logic.

Next, we consider the premise \texttt{ER1}.  This seems, and is
generally considered to be, a reasonable premise for a truly perfect
or greatest being, such as God: if such a being existed, it would
surely not do so merely contingently.  Notice that this premise is
specific to the constant \texttt{g} with the interpretation used here:
the generalization $P \supset \Box P$, where $P$ is a metavariable
ranging over all propositional constants is not a plausible premise,
nor is the instance where \texttt{g} is interpreted as a perfect
island, as in Gaunilo's \citeyearpar{Gaunilo} refutation.

{ \sloppypar The premise \texttt{ER2} is discussed at length in some
philosophical papers \citep[e.g.,][]{Hartshorne61}.  One concern is
that the concept of a perfect or greatest being must be non
self-contradictory (otherwise it could not possibly exist), and this
is not straightforward as there are properties of a perfect being that
seem to conflict (e.g., omnibenevolence and perfect justice).  Another
concern is that non self-contradiction may be insufficient to ensure
the possibility of existence.  However, as we are concerned with logic
and mechanization, we accept \texttt{ER2} as otherwise the proof
cannot proceed.

}

The overall argument seems fairly satisfying: we have one premise
about a hypothesized God's necessary existence, another about his
possible existence, and we are able to conclude his actual existence,
which is something of a surprise and therefore satisfying.

However, we have the background assumption of a symmetric
accessibility relation or the modal axiom B and, in the presence of
this assumption, the premise \texttt{ER1} reduces to $\Diamond g
\supset g$.  This is stated in its general form as \texttt{ER\_triv}
in Figure \ref{er-pvs}; its proof is simply \texttt{(grind\
:polarity?\ t)}.  But now the argument is exposed as trivial:
\begin{center}
$\Diamond g \supset g$, $\Diamond g$, hence $g$.
\end{center}

We will examine several variants on this argument; mostly they
adjust the first premise, sometimes also the conclusion (using $\Box
g$ rather than simply $g$), and sometimes they assume modal axioms
other than B.   But in all cases, we will see that the argument
reduces to the same trivial form.

\subsubsection*{Addendum}$\!$to the published paper (16 August 2020).
If $M$ is the modal assumption, $P_1$ and $P_2$ the premises and $C$
the conclusion, then in classical logic the overall deduction has the
form $M \wedge P_1 \wedge P_2 \vdash C$ and from this it follows that
$M \supset (P_1 \supset (P_2 \supset C))$.  \texttt{ER\_triv} is just
a modal instance of this inevitable relationship.  The reason I label
it trivial is because it is expressed so directly in the premises and
conclusion employed.
\\\textbf{End of Addendum}

\newpage
\subsection{The Version by Kane}
\label{kane}

\cite{Kane84} presents a version of the argument that uses a
slightly more complex formulation of the first premise.  The
interpretation of $g$, the second premise, and the conclusion are the
same as in Eder and Ramharter's version from the previous section.

\begin{description}

\item[Premise K1:] $\Box(g \supset \Box g)$.

Kane reads this as ``Necessarily, if a perfect being exists, then
necessarily a perfect being exists.''  He states that Hartshorne and
others assume that it follows from a principle he names \emph{N}
(not to be confused with the rule of necessitation): ``By
definition, anything which is perfect is such that, if it exists, it
exists necessarily,'' where ``the first $\Box$ in \texttt{K1}
corresponds to the `by definition' of \emph{N}'' \cite[page 337]{Kane84}.

\item[Premise K2:] $\Diamond g$ (i.e., God's existence is possible).

\item[Conclusion KC:] $g$ (i.e., God exists in a classical sense).

\end{description}

The heart of the PVS specification for this version is shown below.
\begin{sessionbiglab}{Kane's version in PVS}
  K1: LEMMA □\,(g => □\,g)

  K_triv: LEMMA symmetric?(access)\,\,IMPLIES □(P => □\,P)\,\,IMPLIES (<>\,P => P)

  K2: AXIOM <>\,g

  KC: THEOREM symmetric?(access) => g
\end{sessionbiglab}
We label the PVS rendition of \texttt{K1} as a \texttt{LEMMA} rather
than an \texttt{AXIOM} because it can be proved from \texttt{ER1}; the
proof is simply \texttt{(grind\ :polarity?\ t)}.  Again, proof of the
conclusion requires modal axiom B or, equivalently, a symmetric
accessibility relation.  Given that \texttt{K1} is entailed by
\texttt{ER1}, it is unsurprising that it also reduces to $\Diamond g
\supset g$, which is stated as \texttt{K\_triv} in the PVS
specification.  Thus, at bottom, this version is based on the same
trivial reasoning as the previous one.

Kane devotes considerable space to discussing the acceptability of
modal axiom B.  But, as explained in the previous section, B is a theorem
of S5, which is the standard formalization for the Alethic logic that
is needed to provide the desired interpretations for $\Box$ and
$\Diamond$, so Kane's agonizing seems otiose.

\subsection{The Version of Malcolm as Reported by Hartshorne}

\citet{Hartshorne61} presents a version of the argument that
he attributes to \citet{Malcolm60}.  This has the same premises
as Eder and Ramharter's version, but the conclusion is $\Box g$ rather
than simply $g$.  Furthermore, rather than modal Axiom B, this version
uses the modal Axiom 5, which is $\Diamond P \supset \Box\Diamond P$.
Axiom 5 corresponds to a Euclidean accessibility relation\footnote{A
relation $R$ is Euclidean if $\forall u, v, w: R(u, v) \wedge R(u, w)
\supset R(v, w)$.} and is assumed in an Alethic logic.

\newpage\noindent
The core of the formal specification in PVS is shown below,
\begin{sessionbiglab}{Malcolm's version in PVS}
  M1: AXIOM g => □\,g

  M_triv: LEMMA Euclidean?(access)\,\,IMPLIES\,\,(P => □\,P)\,\,IMPLIES\,\,(<>\,P => □\,P)

  M2: AXIOM <>\,g

  MC: THEOREM Euclidean?(access) => □\,g

  MC_alt: THEOREM symmetric?(access) => □\,g
\end{sessionbiglab}
PVS easily proves the conclusion \texttt{MC}, but it also proves the
lemma \texttt{M\_triv} and so again the reasoning is too trivial to
claim our interest.

The conclusion can also be proved assuming modal axiom B or,
alternatively, a symmetric accessibility relation, as stated in
\texttt{MC\_alt}: we first use the method of Eder and Ramharter's
version to establish $g$, then a second application of the premise
\texttt{M1} delivers $\Box g$.

\subsection{The Version of Adams as Reported by Matthews}
\label{adams}

\citet{Adams71} derives a modal argument, not from \emph{Proslogion}
Chapter III, but from Anselm's reply \citeyearpar[Chapter I, paragraph
6]{Anselm:reply} to Gaunilo \citeyearpar{Gaunilo}.  Despite
its different origin, this argument is the same as Kane's.  However,
\citet{Matthews:ontological}, who reports Adam's version, gives the
conclusion as $\Box g$.  I will use this latter version as it
completes the set of permutations on the first premise and conclusion.

This version of the argument in PVS is shown below.
\begin{sessionbiglab}{Adams' Version in PVS}
  A1: LEMMA □(g => □\,g)

  A_triv: LEMMA
      Euclidean?(access) IMPLIES □(P => □\,P) IMPLIES (<>\,P => □\,P)

  A2: AXIOM <>\,g

  AC: THEOREM Euclidean?(access) => □\,g

  AC_alt: THEOREM symmetric?(access) => □\,g
\end{sessionbiglab}

As we expect, based on the other variants we have examined, this
version can be proved assuming either a symmetric or Euclidean
accessibility relation (i.e., standard axioms B or 5, respectively).
And with a Euclidean accessibility relation, the lemma
\texttt{A\_triv} reveals that Premise \texttt{A1} reduces to
$\Diamond g \supset \Box g$ so the conclusion follows trivially from
Premise \texttt{A2}.

\newpage
\subsection{The Version by Hartshorne}
\label{hart1}

\citet[pp.\ 50--51]{Hartshorne:perfection} presented the
first formal reconstruction of the modal argument.  This uses a
different and more complex formulation of the first premise than the
others we have examined.  The interpretation of $g$, the second
premise, and the conclusion are the same as in the versions of Eder
and Ramharter, and Kane.

\begin{description}

\item[Premise H1:] $ g \strictif \Box g$, where $\strictif$
denotes \emph{strict} implication.

\item[Premise H2:] $\Diamond g$ (i.e., God's existence is possible).

\item[Conclusion HC:] $g$ (i.e., God exists in a classical sense).

\end{description}

We say that $P$ \emph{strictly} implies $Q$ if it is not possible for $P$ to
be true and $Q$ false: that is, in an Alethic logic $$ P \strictif Q
\defn \neg \Diamond (P \wedge \neg Q).$$ It is a theorem that strict
implication is the same as necessary material implication: $$ P
\strictif Q = \Box (P \supset Q).$$ This equality is a theorem of all
modal logics (i.e., it requires no axioms) but it carries the intended
interpretation only in Alethic logics.  Thus, premise \texttt{H1} is
equivalent to Kane's premise \texttt{K1} of Section \ref{kane} and the
rest of the argument is also the same as Kane's.  It follows that the
conclusion requires the assumption of a symmetric accessibility
relation or, equivalently, modal axiom B, and that under this
assumption the premise \texttt{H1} reduces to $\Diamond g \supset g$
and the argument is thereby seen to be trivial.

In PVS, this is written as follows, where we
use \texttt{|>} for strict implication \citep{Rushby:Modal17}.
\begin{sessionbiglab}{Hartshorne's version in PVS}
  |>(P, Q): pmlformulas =  \wig\,<>(P & \wig\,Q)

  H1: AXIOM g |> □\,g

  H_triv: LEMMA symmetric?(access) IMPLIES (P |> □\,P) IMPLIES (<>\,P => P)

  H2: AXIOM <>\,g

  HC: THEOREM symmetric?(access) => g
\end{sessionbiglab}
The conclusion \texttt{HC} can be proved by the following PVS
commands, whose length is due to weak quantifier reasoning in PVS.
\begin{sessionlab}{PVS proof}
(lemma "H1") (lemma "H2")
(grind :if-match nil)
(inst?) (skosimp) (inst?)
(grind :polarity? t)
\end{sessionlab}
Similarly, \texttt{H\_triv} is proved as follows.
\begin{sessionlab}{PVS proof}
(grind :if-match nil) (inst? -2) (grind :polarity? t)
\end{sessionlab}

We know from Section \ref{adams} that Adams' version is the same
as Kane's with the conclusion changed to $\Box g$ and the modal
assumption changed to a Euclidean accessibility relation or,
equivalently, modal axiom 5.  Since Hartshorne's version is equivalent
to Kane's we can perform the same transformations here to yield the
following variant.
\begin{sessionbiglab}{Alternative Hartshorne PVS}
  H_triv_alt: LEMMA
     Euclidean?(access) IMPLIES (P |> □\,P) IMPLIES (<>\,P => □\,P)

  HC_alt: THEOREM Euclidean?(access) IMPLIES □\,g
\end{sessionbiglab}
These are proved in the same way as the originals.

\vspace*{-1ex}
\subsubsection*{Addendum}$\!$to the published paper (29 July 2021).
Harold Thimbleby remarked that it is a pity that PVS imposes an ASCII
rendition and does not reproduce mathematical notation.  I thank
Harold for prompting me to report that, in fact, recent versions of
PVS do support extended character sets with mathematical symbols, and
all versions of PVS are able to typeset specifications in \LaTeX.  A
\LaTeX-printed version of Hartshorne's version is shown below.

\def\munderscoretimestwofn#1#2{{#1 \times #2}}
\def\fmunderscoretimestwofn#1#2{{#1 \times #2}}
\def\sigmaunderscoretimestwofn#1#2{{#1 \times #2}}
\def\generatedunderscoresubsetunderscorealgebraonefn#1{{{\cal A}(#1)}}
\def\generatedunderscoresigmaunderscorealgebraonefn#1{{{\cal S}(#1)}}
\def\aeunderscoredecreasingotheronefn#1{{\pvsid{decreasing?}(#1)~\mbox{\it a.e.}}}
\def\aeunderscoreincreasingotheronefn#1{{\pvsid{increasing?}(#1)~\mbox{\it a.e.}}}
\def\aeunderscoreconvergenceothertwofn#1#2{{#1 \longrightarrow #2~\mbox{\it a.e.}}}
\def\aeunderscoreeqothertwofn#1#2{{#1 = #2~\mbox{\it a.e.}}}
\def\aeunderscoreleothertwofn#1#2{{#1 \leq #2~\mbox{\it a.e.}}}
\def\aeunderscoreposotheronefn#1{{#1> 0~\mbox{\it a.e.}}}
\def\aeunderscorenonnegotheronefn#1{{#1 \geq 0~\mbox{\it a.e.}}}
\def\aeunderscorezerootheronefn#1{{#1 = 0~\mbox{\it a.e.}}}
\def\xunderscorelttwofn#1#2{{#1 < #2}}
\def\xunderscoreletwofn#1#2{{#1 \leq #2}}
\def\xunderscoreeqtwofn#1#2{{#1 = #2}}
\def\xunderscoretimestwofn#1#2{{#1 \times #2}}
\def\xunderscoreaddtwofn#1#2{{#1 + #2}}
\def\xunderscorelimitonefn#1{{\pvsid{limit}(#1)}}
\def\xunderscoresumonefn#1{{\sum #1}}
\def\xunderscoresigmathreefn#1#2#3{{\sum_{#1}^{#2} #3}}
\def\xunderscoresuponefn#1{{\pvsid{sup}(#1)}}
\def\xunderscoreinfonefn#1{{\pvsid{inf}(#1)}}
\def\pointwiseunderscoreconvergesunderscoredowntoothertwofn#1#2{{#1 \searrow #2}}
\def\pointwiseunderscoreconvergesunderscoreuptoothertwofn#1#2{{#1 \nearrow #2}}
\def\pointwiseunderscoreconvergenceothertwofn#1#2{{#1 \longrightarrow #2}}
\def\convergesunderscoredowntoothertwofn#1#2{{#1 \searrow #2}}
\def\convergesunderscoreuptoothertwofn#1#2{{#1 \nearrow #2}}
\def\convergenceothertwofn#1#2{{#1 \longrightarrow #2}}
\def\convergencetwofn#1#2{{#1 \longrightarrow #2}}
\def\crossunderscoreproducttwofn#1#2{{#1 \times #2}}
\def\conjugateonefn#1{{\overline{#1}}}
\def\cunderscoredivtwofn#1#2{{#1/#2}}
\def\cunderscoremultwofn#1#2{{#1\times#2}}
\def\cunderscoresubtwofn#1#2{{#1-#2}}
\def\cunderscorenegonefn#1{{-#1}}
\def\cunderscoreaddtwofn#1#2{{#1+#2}}
\def\Imonefn#1{{\Im(#1)}}
\def\Reonefn#1{{\Re(#1)}}
\def\Etwofn#1#2{{\mathbb{E}(#1~|~#2)}}
\def\Eonefn#1{{\mathbb{E}(#1)}}
\def\Ptwofn#1#2{{\mathbb{P}(#1~|~#2)}}
\def\Ponefn#1{{\mathbb{P}(#1)}}
\def\xtwofn#1#2{{#1\times#2}}
\def\asttwofn#1#2{{#1\ast#2}}
\def\minusonefn#1{{{#1}^{-}}}
\def\plusonefn#1{{{#1}^{+}}}
\def\astonefn#1{{{#1}^{\ast}}}
\def\dottwofn#1#2{{#1\bullet#2}}
\def\integralthreefn#1#2#3{{\int_{#1}^{#2} #3}}
\def\integraltwofn#1#2{{\int_{#1} #2}}
\def\integralonefn#1{{\int#1}}
\def\normonefn#1{{\left||{#1}\right||}}
\def\phionefn#1{{\pvssubscript{\phi}{#1}}}
\def\infunderscoreclosedonefn#1{{\left(-\infty,~#1\right]}}
\def\closedunderscoreinfonefn#1{{\left[#1,~\infty\right)}}
\def\infunderscoreopenonefn#1{(-\infty,~#1)}
\def\openunderscoreinfonefn#1{(#1,~\infty)}
\def\closedtwofn#1#2{{\left[#1,~#2\right]}}
\def\opentwofn#1#2{(#1,~#2)}
\def\sigmathreefn#1#2#3{{\sum_{#1}^{#2} #3}}
\def\sigmatwofn#1#2{{\sum_{#1} {#2}}}
\def\ceilingonefn#1{{\lceil{#1}\rceil}}
\def\flooronefn#1{{\lfloor{#1}\rfloor}}
\def\absonefn#1{{\left|{#1}\right|}}
\def\roottwofn#1#2{{\sqrt[#2]{#1}}}
\def\sqrtonefn#1{{\sqrt{#1}}}
\def\sqonefn#1{{\pvssuperscript{#1}{2}}}
\def\expttwofn#1#2{{\pvssuperscript{#1}{#2}}}
\def\opcarettwofn#1#2{{\pvssuperscript{#1}{#2}}}
\def\indexedunderscoresetsotherIIntersectiononefn#1{{\bigcap #1}}
\def\indexedunderscoresetsotherIUniononefn#1{{\bigcup #1}}
\def\setsotherIntersectiononefn#1{{\bigcap #1}}
\def\setsotherUniononefn#1{{\bigcup #1}}
\def\setsotherremovetwofn#1#2{{(#2 \setminus \{#1\})}}
\def\setsotheraddtwofn#1#2{{(#2 \cup \{#1\})}}
\def\setsotherdifferencetwofn#1#2{{(#1 \setminus #2)}}
\def\setsothercomplementonefn#1{{\overline{#1}}}
\def\setsotherintersectiontwofn#1#2{{(#1 \cap #2)}}
\def\setsotheruniontwofn#1#2{{(#1 \cup #2)}}
\def\setsotherstrictunderscoresubsetothertwofn#1#2{{(#1 \subset #2)}}
\def\setsothersubsetothertwofn#1#2{{(#1 \subseteq #2)}}
\def\setsothermembertwofn#1#2{{(#1 \in #2)}}
\def\opohtwofn#1#2{{#1\circ#2}}
\def\opdividetwofn#1#2{{#1 / #2}}
\def\optimestwofn#1#2{{#1\times#2}}
\def\opdifferenceonefn#1{{-#1}}
\def\opdifferencetwofn#1#2{{#1-#2}}
\def\opplustwofn#1#2{{#1+#2}}

\def\optrianglerighttwofn#1#2{{#1 \strictif #2}}
\def\pvsid#1{\textrm{#1}}
\def\pvsdeclspacing{0in}		

\def\pvskey#1{\textbf{\uppercase{#1}}}

\begin{session}
  \(\optrianglerighttwofn{P}{Q}\): \pvsid{pmlformulas} \(\defn\) \(\neg\)\(\Diamond\)\pvsid{(}\(P\) \(\wedge\) \(\neg\)\(Q\)\pvsid{)}\vspace*{\pvsdeclspacing}

  \(\pvsid{H1}\): \pvskey{AXIOM} \(\optrianglerighttwofn{g}{\Box{}g}\)\vspace*{\pvsdeclspacing}

  \(\pvsid{H1\_triv}\): \pvskey{LEMMA} \pvsid{symmetric?}\pvsid{(}\pvsid{access}\pvsid{)} \(\supset\) \((\optrianglerighttwofn{P}{\Box{}P})\) \(\supset\) \((\Diamond\)\(P\) \(\supset\) \(P)\)\vspace*{\pvsdeclspacing}

  \(\pvsid{H2}\): \pvskey{AXIOM} \(\Diamond\)\(g\)\vspace*{\pvsdeclspacing}

  \pvsid{HC}: \pvskey{THEOREM} \pvsid{symmetric?}\pvsid{(}\pvsid{access}\pvsid{)} \(\supset\) \(g\)\vspace*{\pvsdeclspacing}
\end{session}

Exactly how PVS ASCII text is rendered in \LaTeX\ is controlled by a
``substitutions'' file \texttt{pvs-tex.sub}.  A comprehensive
substitution file is provided with PVS but it can be augmented by the
user.  The rendition above was generated using the following
substitutions file as augmentation.  The first column in this file
identifies the ASCII source to be substituted, the second identifies
the kind of PVS object it is (see the PVS documentation), the third
gives the size of the substitution in \emph{em}s, and the final column
gives the desired \LaTeX\ substitution.

\begin{sessionbiglab}{PVS \LaTeX\ substitution file}
\begin{verbatim}
|>               2      3       {#1 \strictif #2}
|>              id      1       \strictif
H1              id      2       \pvsid{H1}
H1_triv         id      3       \pvsid{H1\_triv}
H2              id      2       \pvsid{H2}
IMPLIES         id      3       \supset
=               key     2       \defn
~               id      1       \sim
~               id      1       \neg
=>              id      1       \supset
&               id      1       \wedge
\end{verbatim}
\end{sessionbiglab}

It might be considered that these substitutions are too aggressive
because they do not distinguish connectives to be
interpreted modally from those that are propositional.   Removing
the last three lines from the substitution file above (where the
third-last line overrides the fourth-last), restores this
distinction and generates the following rendition.

\begin{session}
  \(\optrianglerighttwofn{P}{Q}\): \pvsid{pmlformulas} \(\defn\) \(\sim\)\(\Diamond\)\pvsid{(}\(P\) \(\&\) \(\sim\)\(Q\)\pvsid{)}\vspace*{\pvsdeclspacing}

  \(\pvsid{H1}\): \pvskey{AXIOM} \(\optrianglerighttwofn{g}{\Box{}g}\)\vspace*{\pvsdeclspacing}

  \(\pvsid{H1\_triv}\): \pvskey{LEMMA} \pvsid{symmetric?}\pvsid{(}\pvsid{access}\pvsid{)} \(\supset\) \((\optrianglerighttwofn{P}{\Box{}P})\) \(\supset\) \((\Diamond\)\(P\) \(\Rightarrow\) \(P)\)\vspace*{\pvsdeclspacing}

  \(\pvsid{H2}\): \pvskey{AXIOM} \(\Diamond\)\(g\)\vspace*{\pvsdeclspacing}

  \pvsid{HC}: \pvskey{THEOREM} \pvsid{symmetric?}\pvsid{(}\pvsid{access}\pvsid{)} \(\Rightarrow\) \(g\)\vspace*{\pvsdeclspacing}
\end{session}
\vspace*{-2ex}
\textbf{End of Addendum}

\vspace*{-2ex}
\section{The Utility of These Methods}
\label{discuss}

In my view, the utility of the methods illustrated here, and of
computational philosophy more generally, is not so much that they
enable verification of hard logical problems (those abound in logic
and mathematics, not so much in philosophy) but that they enable rapid
and error-free exploration of multiple formulations of the same or
similar problems.  This allows discovery of equivalences and subtle
differences, not to mention checking of validity and discovery of
economical and persuasive formulations.

In the case of the five specific formulations of the Modal Ontological
Argument examined in Section \ref{modal-arg}, we found that all of
them are variations on a simple pattern.  Their authors and subsequent
commentators did not seem to recognize this similarity, nor some of
the differences within it.  Hartshorne's
\citeyearpar{Hartshorne:perfection} formulation uses strict
implication whereas his earlier \citeyearpar{Hartshorne61} treatment
of Malcolm's \citeyearpar{Malcolm60} original exposition does not.
\citet[Section 4.1]{Eder&Ramharter15} state that they follow
Hartshorne's \citeyearpar{Hartshorne:perfection} formulation, but then
use material rather than strict implication.  \cite{Kane84} states
that his formulation is from \citet{Hartshorne:perfection} but
``readers of that work and subsequent journal literature on the modal
OA may not recognize the adaptation.''  He attributes it to an
unpublished simplification due to C. Anthony Anderson
\cite[p. 339]{Kane84}.  \citet{Adams71} treatment is derived not from
the \emph{Proslogion}, but from Anselm's reply to Gaunilo.

Despite the apparent differences in their original presentations, all
of the reconstructions have a first premise that asserts, in one way
or another, that if a perfect or greatest being (i.e., God) exists,
then he exists necessarily (i.e., $g \supset \Box g$, $\Box(g \supset
\Box g$), or $g \strictif \Box g$); a second premise that asserts his
existence is possible (i.e., $\Diamond g$); and a conclusion that he
exists or necessarily exists (i.e., $g$ or $\Box g$).

Proof that the conclusion follows from the premises requires certain
properties of the modal logic concerned; these can be expressed either
by citing one of the standard modal axioms (i.e., B or 5) or the
corresponding property of the accessibility relation on possible
worlds (i.e., symmetric or Euclidean, respectively).  The argument
requires an Alethic logic (one where the modalities express necessity
and possibility) and these have accessibility relations that are
equivalence relations or, alternatively, assume modal axioms T and
5, and have B as a theorem; consequently, there can be no objection to
the modal properties required in the proofs.

In addition to verifying the different formulations of the argument,
we showed that in each case, the modal axiom that was assumed is
sufficient to prove that the first premise entails a formula that
states that the second premise directly entails the conclusion, that
is: $$\emph{Modal axiom} \supset (\emph{Premise 1} \supset
(\emph{Premise 2} \supset \emph{Conclusion})).$$ Thus, the reasoning is
essentially trivial.

Mechanization is needed in this kind of examination for the same
reasons that automated calculation is used in other branches of
engineering and science: it is simply too tedious and too error-prone
to conduct repetitive analyses by hand.  Some may concede this for
hard problems but dismiss it for the Modal Ontological Argument: one
reviewer wrote ``the results of the analysis have been common
knowledge in the field for at least the past forty years (at least
among those with even minimal competence in propositional modal
logic).''  I would respectfully disagree that these results have been
common knowledge (see the second paragraph of this section) and would
also be cautious about widespread
reliability, if not competence, in modal logic.  To quote \citet[p.\
175]{Lewis70}:
\begin{quote}
``Philosophy abounds in troublesome modal arguments---endlessly
debated, perennially plausible, perennially suspect. The standards of
validity for modal reasoning have long been unclear; they become clear
only when we provide a semantic analysis of modal logic by reference
to possible worlds\ldots''
\end{quote}

For illustration, this journal published a paper by
\citet{Jacquette97} in which he criticizes Hartshorne's
\citeyearpar{Hartshorne:perfection} formulation of the Modal
Ontological Argument.  Jacquette presents a semi-formal 10-line proof
that he attributes to Hartshorne and finds it defective on several
grounds.  Most damagingly, he claims there is a
\begin{quote}
``logical difficulty in a key assumption of the proof that renders the
entire inference unsound.  Proposition (5), which Hartshorne says
follows from (1) as a modal form of \emph{modus tollens}, is clearly
false.''  
\end{quote}
The proposition Jacquette refers to as (1) is Anselm's Principle $g
\supset \Box g$ and (5) 
is $\Box\neg\Box g \supset \Box\neg{}g$ which, contrary to Jacquette,
is a perfectly valid deduction from (1), as can be verified by adding
it to any of the PVS analyses presented in Section
\ref{modal-arg} and using similar proof steps.

Jacquette attributes what he considers Hartshorne's error to his use
of a ``modal form of \emph{modus tollens}'' that Jacquette writes as
$(\alpha \rightarrow \beta) \rightarrow (\Box\sim\beta\ \rightarrow\
\Box\sim\alpha)$ and states ``is not generally true.''  He presents a
counterexample in which $\alpha$ is ``snow is red'' and $\beta$ is
``2+2 = 5.''  This single formula highlights a couple of interesting
pitfalls in semi-formal use of modal logic.  First of all, Hartshorne
uses \emph{strict} implication, which we write as $\strictif$ to
distinguish it from the $\supset$ of material implication.  Jacquette
makes no mention of this and it is not clear whether his $\rightarrow$
is intended to represent strict or material implication.  

Second, it is often necessary to be careful about which parts of a
sentence are to be interpreted modally, and which are conventional
propositional logic.  Specifically, a correct statement of modal
\emph{modus tollens} is expressed in PVS as follows (i.e., two modal
sentences connected propositionally),
\begin{sessionlab}{PVS text}
  tollens: LEMMA (P => Q) IMPLIES (□\wig{}Q => □\wig{}P)
\end{sessionlab}
whereas the following (i.e., a single modal sentence) is invalid.
\begin{sessionlab}{PVS text}
  tollens_bad: CLAIM (P => Q) => (□\wig{}Q => □\wig{}P)
\end{sessionlab}
Jacquette seems to be using the invalid form, but we cannot be sure
because his notation lacks the necessary distinctions.  The key
difference is that the correct form quantifies possible worlds
separately for each side of the implication, whereas the invalid form
requires the whole sentence to be true in the same worlds.  This is
similar to the reason that the deduction theorem is generally invalid
in modal logic.

These errors in Jacquette's analysis are easily detected using truly
formal and mechanically checked specifications.  Furthermore, we or
Jacquette could easily satisfy ourselves that even if we have qualms
about Hartshorne's proof (i.e., selection of inference steps), his
formulation of the argument (i.e., selection of premises and
conclusion) is valid.

Although Jacquette's charge of invalidity for Hartshorne's formulation
of the argument is mistaken, he raises other points that have
merit.  In particular, he observes that Anselm's presentation has the
form of a \emph{reductio ad absurdum} whereas Hartshorne's (and we
might add, all others we have considered) does not.  He also notes
that Anselm uses the notion of ``greater than'' (i.e., an ordering) on
beings and that he uses ``conceivable'' as a modality that seems
different than the ``possible'' of alethic logic.  These points are
all abstracted away in conventional formalizations of the argument.
Accordingly, Jacquette \citeyearpar[Section 4]{Jacquette97} uses ideas that he
attributes to \citet{Priest2002} to construct a formalization of the
argument that remedies these deficiencies.  His treatment begins as
follows.

\begin{description}
\item[1.][Definition of $g$:] $g = \delta x: \neg \exists y: \Diamond (y >
x)$.\\ Here, $\delta$ is ``indifferently a definite or indefinite
descriptor'' (i.e., ``the'' or ``an'') and I have substituted
$\Diamond$ for Jacquette's $\tau$, which he intends as ``an
intensional modality'' (since no semantics are given, this shift to a
more familiar notation has no impact).
\end{description}

An informal reading is that God is some being than which no greater is
conceivable.  The construction $\neg \exists y: \Diamond (y > x)$ is
also used in formalizations using first order modal logic for the
traditional, Proslogion Chapter II, Ontological Argument \cite[Section
4.2]{Eder&Ramharter15}.  These modal formalizations of the traditional
argument have been mechanized in PVS \cite[Section
3.1]{Rushby:modalont19}, as has a version in classical first-order
logic that uses definite descriptions
\citep{Rushby:ontological13},
where indefinite descriptions
or choice functions, and Hilbert's $\varepsilon$, are also discussed.

We might wonder whether the given construction is
equivalent to $\neg \Diamond \exists y: (y > x)$ (i.e., interchange
the modality and quantifier) and the answer is: it depends on whether
we have constant or varying domains and, if varying, whether they are
nonincreasing or nondecreasing, thereby highlighting some of the
subtlety lurking here \citep[Section 3]{Rushby:Modal17}.

The next interesting line in Jacquette's treatment is the following.
\begin{description}
\item[3.] [Conceivable greatness:] $\forall x: (\neg E!x \supset \exists y:
\Diamond (y > x))$.\\
Here, $E!x$ is a predicate indicating the ``real existence'' of $x$.
\end{description}
Jacquette uses the definition 1 and premise 3 in a \emph{reductio}
proof that establishes $E!g$: that is, the real existence of God.

Now, we could attempt to verify his proof in PVS (in fact, I have done
this) but I am reluctant, this late in the paper, to extend
description of the PVS embedding of modal logic from propositional to
first order and to provide a fully formal presentation, so I will make
my point using Jacquette's informal
notation, but the following derives from experience applying first
order modal reasoning in PVS to formal verification of the Proslogion
Chapter II argument, where the methods are fully described
\citep{Rushby:modalont19}.

First, Jacquette does not indicate whether $g$ and $E!x$ are to be
interpreted flexibly (i.e., dependent on the world) or rigidly (i.e.,
independent of it).  I assume a rigid interpretation.  Then, nowhere
in Jacquette's proof is the construction $\exists y: \Diamond (y > x)$
opened up to reveal either $y$ or the $>$ relation.  Thus, we can
replace the whole thing by an uninterpreted function $F(x)$.  Then 1
and 3 above become the following ``primed'' variants;
\begin{description}
\item[1$'$:] $g = \delta x: \neg F(x) $,

\item[3$'$:] $\forall x: \neg E!x \supset F(x)$,
\end{description}
and we can replace 3$'$ by its contrapositive
\begin{description}
\item[3$''$:] $\forall x: \neg F(x) \supset E!x$.
\end{description}
But now the deduction is revealed as trivial: we have $\neg F(g)$ from
1$'$ and instantiate 3$''$ to give $\neg F(g) \supset E!g$ and the
conclusion $E!g$ is immediate.  Furthermore, the proof has lost all
trace of modal reasoning.  Jacquette's goal was to make explicit more
of Anselm's reasoning, but he ends up trivializing it.

Although his execution is defective, we can agree with Jacquette's
motivation that a faithful reconstruction of Anselm's argument should
represent his language explicitly.  A reviewer for this paper makes
the same point and suggests that PVS could be used systematically to
explore different renditions for ``conceive,'' ``nothing greater''
etc.  I am sympathetic to this but suspect that the defective
renditions would exhibit triviality in the style of Jacquette's
example rather than invalidity, and this is more difficult to detect
automatically.  I hope others may wish to explore some of the
alternatives and that verification tools such as PVS will assist their
endeavors.

\section{Conclusion}
\label{conc}

We have examined several reconstructions of the modal ontological
argument from Anselm's Proslogion Chapter III and his reply to
Gaunilo.  When fully formalized, all of the reconstructions have a
very similar form, and are proved in the same way.

The proofs require certain properties of the modal logic employed,
and these can be specified either by citing one of the standard modal
axioms (i.e., B or 5), or the corresponding property of the
accessibility relation on possible worlds (i.e., symmetric or
Euclidean, respectively).  The argument requires an Alethic logic (one
where the modalities express necessity and possibility) and these have
accessibility relations that are equivalence relations or,
alternatively, assume modal axioms T and 5, and have B as a theorem;
consequently, there can be no objection to the modal properties
required in the proofs.

We expressed the various reconstructions in PVS and mechanically
verified their validity.  We also showed that in each case, the modal
axiom that was assumed is sufficient to prove that the first premise
entails a formula that states that the second premise directly entails
the conclusion, that is: $$\emph{Modal axiom} \supset (\emph{Premise 1}
\supset (\emph{Premise 2} \supset \emph{Conclusion})).\footnote{This
relationship is inevitable in a deduction with only two premises.
What is surprising, and the reason I label it trivial, is how directly
it is expressed by the premises employed.}$$  Thus, the
reasoning is essentially trivial and standard readings of the modal
ontological argument should not retain our interest.

We also showed how mechanically assisted formalization and analysis
could expose flawed treatments, including one published in this
journal, and also vacuous formalizations that present the appearance of
Anselm's argument but are logically trivial.

\citet{Campbell18:book,Campbell19:AnselmP3} provides an alternative
reading in which Chapter III of the Proslogion does not stand alone
but continues the argument of Chapter II.  Campbell interprets Anselm
differently than the scholars considered here and constructs a
different translation into English.  Formal analysis of his and
related readings will require separate treatment.

This paper is the fifth is a series where I apply mechanized
verification to analysis of formal reconstructions of Anselm's
Ontological arguments.  In the first \citep{Rushby:ontological13},  I
examined a formalization by \cite{Oppenheimer&Zalta11} for the
traditional Proslogion II argument that uses definite descriptions.
I showed that this formalization is very close to circular (i.e., begs
the question) in an informal sense.  In the second
\citep{Rushby18:begging}, I gave formal definitions for question
begging in formal proofs and I showed that all the examined
reconstructions of the Proslogion II argument in classical (i.e.,
first- and higher-order) logic were guilty of begging the question.
In a revision and extension to that paper
\citep{Rushby19:ontargbegsvac}, I additionally showed that all these
reconstructions entail variants that apply no interpretation to
``something than which there is no greater'' and are therefore vacuous
and vulnerable to Gaunilo's \citeyearpar{Gaunilo} refutation, and I
argued that the basic reconstructions inherit these charges.  In the
fourth \citep{Rushby:modalont19}, I extended these analyses to
reconstructions of the Proslogion II argument in quantified modal
logic and found they suffer the same defects.  In the present paper,
we have examined reconstructions of the Proslogion III argument in
propositional modal logic and found it to be trivial once the assumed
modal axiom is taken into account.

Mechanization is necessary for these analyses because they require
detailed scrutiny of small variations in premises and assumptions.
This is tedious and error-prone to do by hand, but fast, simple, and
even enjoyable to accomplish with mechanized assistance.  I hope these
exercises may encourage others to apply modern formal verification
tools to examination of other suitable philosophical and theological
topics.


\subsubsection*{Acknowledgements.}
The challenging comments of a reviewer helped me improve the paper.  I
am also grateful to my colleagues St\'{e}phane Graham-Lengrand and
N. Shankar for discussions on modal logic and its automation.
This work was funded by SRI International and by my retirement plan.


\renewcommand\bibname{foobar}
\renewcommand\bibsection{\section*{\refname}}
\bibliographystyle{unsrtnat}

\appendix
\newpage
\section*{Postscript}

This paper was published in its original form in April 2021; at almost
the same time, Andrzej Bi{\l}at published a paper on related topics
\citep{Bilat21}.  His purpose was to identify some alternative premises
that can formalize the Modal Ontological Argument using fewer (or no)
modal axioms.  His alternative premises replace those traditionally
used to formalize Anselm's Principle (recall Section \ref{er}) and,
because they require fewer modal axioms, he considers the arguments
corresponding to his alternative premises to be simpler than others,
including those presented here in Section \ref{modal-arg}.

Much of his paper is concerned with defining and demonstrating
``relevance'' of the various premises and I have nothing to add to
that discussion.  I do, however, wish to note that his alternative
premises can be derived by applying the discarded modal axioms to a
standard (e.g., Kane's) formalization of Anselm's Principle.  Thus,
the alternative premises support a proof with fewer modal axioms
simply because the discarded modal axioms are already ``built in'' to
the alternative premises: those premises are, in effect, lemmas that
are proved using the discarded axioms.  Furthermore, the alternative
premises are far less plausible than the standard ones and they
trivialize the argument.

Before getting into details of the specific premises, I would like to
challenge the motivation for seeking fewer or ``simpler'' modal
axioms.  In order for formalizations of the Modal Ontological Argument
to carry the intended interpretation, the modal qualifiers must
represent ``necessary'' and ``possible,'' which means we need an
Alethic Logic, and it is generally accepted that these correspond to
logics with Modal Axioms 5 and T (i.e., S5), which have B as a
theorem.  Given that these modal axioms are part of the required
logical infrastructure, I see no harm in using them freely.

Bi{\l}at starts with the premises $\Diamond g$ and $\sim\!\Diamond g \
\vee\ \Box g$, which he names (1) and (2) respectively and from which he
derives the conclusion $g$ using Modal Axiom T.\footnote{Bi{\l}at uses
M for $\Diamond$ and L for $\Box$.}  We note that (2) is equivalent to
$\Diamond g \supset \Box g$ and so $\Box g$ follows immediately from (1)
and then Modal Axiom T reduces this to $g$.

Bi{\l}at observes that the argument using (2) is simpler than those
using traditional forms of Anselm's Principle such as $\Box (g \supset
\Box g)$ (as in Kane's treatment), which he names (3), because it
requires fewer or simpler modal axioms (i.e., T as opposed to 5 or
B)\footnote{And would require none if we accepted $\Box g$ as the
conclusion.}.

This postscript was motivated by the observation that this is a
somewhat contrived result because (2) is a lemma that follows from
applying Modal Axiom 5 to (3).  We demonstrated this (for Hartshorne'
version of Anselm's Principle) in lemma
\texttt{H\_triv\_alt} in Section \ref{hart1}.  We provide a more
focused demonstration in the PVS theory shown in Figure
\ref{bilat-pvs}.

Here, axioms \texttt{B1}, \texttt{B2}, and \texttt{B3} correspond to
Bi{\l}at's (1), (2), and (3), respectively.  \texttt{B2triv} proves
that \texttt{B2} follows from \texttt{B3}, given a Euclidean access
relation (which corresponds to Modal Axiom 5), and BC2 proves the
conclusion \texttt{g} from \texttt{B1} and \texttt{B2}, given a
reflexive access relation (which corresponds to Modal Axiom T).  The
PVS proofs are essentially similar to those shown in Section \ref{modal-arg}.

\begin{figure}[ht]
\begin{sessionlab}{PVS Theory}
Bilat: THEORY
BEGIN

IMPORTING modal_axioms

  g: pvars

  B1: AXIOM <> g

  B2: AXIOM \wig <> g \verb|\|/ □ g

  B3: AXIOM □ (g => □ g) 

  B2triv: LEMMA B3 AND Euclidean?(access) IMPLIES B2

  BC2: THEOREM B1 AND B2 AND reflexive?(access) IMPLIES g

  B4: AXIOM \wig{}g => \wig{} <> g

  B4triv: LEMMA B2 AND reflexive?(access) IMPLIES B4

  BC4: THEOREM B1 AND B4 IMPLIES g

END Bilat
\end{sessionlab}
\caption{\label{bilat-pvs}Bi{\l}at's Alternative Premises in PVS}
\end{figure}

Bi{\l}at next introduces the formula $\sim\! g\ \supset\ \sim\Diamond
g$, which he names (4), and remarks that this can be used to prove the
conclusion $g$ with no modal axioms.  Again this is hardly surprising
as (4) is equivalent to (its contrapositive) $\Diamond g \supset g$
and the conclusion $g$ then follows immediately from (1).  The PVS
theorem \texttt{BC4} confirms this.

Since it uses no modal axioms, Bi{\l}at claims that the argument with
(4) is simpler than that with (2).  Again, we assert this is contrived
because (4) is the result of applying Modal Axiom T (which is
equivalent to a reflexive access relation) to (2), as stated and
proved in the PVS formula \texttt{B4triv}.\footnote{Alternatively, it
is also the result of applying Modal Axiom B (symmetry of the access
relation) to (3) as demonstrated in PVS \texttt{LEMMA}
\texttt{K\_triv} in Section \ref{kane}.  Bi{\l}at himself notes this
\cite[Metatheorem 7(ii)]{Bilat21}.  He also states that (4) is
directly entailed by (2) \cite[Metatheorem 7(i)]{Bilat21}, but this is
incorrect: it requires Modal Axiom T.}

We noted in Section \ref{conc} that the overall form of the
modal argument is $$\emph{Modal Axioms} \supset (\emph{Premise 1}
\supset (\emph{Premise 2} \supset \emph{Conclusion}))$$ or
equivalently $$(\emph{Modal Axioms} \wedge \emph{Premise 1}) \supset
(\emph{Premise 2} \supset \emph{Conclusion}),$$ where \emph{Modal
Axioms} may be a conjunction of axioms.

Clearly, we can nominate some fewer or otherwise ``simpler''
\emph{Modal Axioms}$'$ (which may be empty) by discarding or replacing
some of the \emph{Modal Axioms} and can then invent a \emph{Premise\
1}$'$ such that $$(\emph{Modal Axioms} \wedge \emph{Premise 1})
\supset (\emph{Modal Axioms}' \wedge \emph{Premise 1}')$$ and
$$(\emph{Modal Axioms}' \wedge \emph{Premise 1}') \supset
(\emph{Premise 2} \supset \emph{Conclusion}).$$ and then claim that
the second argument is simpler than the first because it uses fewer or
simpler modal axioms.

Similarly, we may also be able to invent \emph{Premise 1}$''$ such
that $$(\emph{Modal Axioms} \wedge \emph{Premise 1})
\supset \emph{Premise 1}'' $$ and
$$\emph{Premise 1}'' \supset (\emph{Premise 2} \supset \emph{Conclusion})$$
and can declare that this argument is yet simpler than the first as it
requires no modal axioms at all.  But in each case, the new premise is simply
the result of applying the previous modal axioms to the previous
premise, so the new premise ``builds in'' the discarded modal axioms:
they are not independent premises but lemmas.

Furthermore, the alternative premises reduce the argument to
triviality.  The standard formulations of Anselm's Principle, such as
$g \supset \Box g$, $\Box(g \supset \Box g)$, or $g \strictif \Box g$
quite plausibly express the intuition that if God exists, then his
existence cannot be contingent (i.e., chance) but must be necessary.
Given any one of these, it then requires (1) plus modal axioms T and 5
or B to deduce his actual existence.  As already noted, the argument
requires an Alethic logic (i.e., one where the modalities express
necessity and possibility) and it is generally accepted that this
requires modal axioms T and 5 (i.e., S5), from which B follows as a
theorem (equivalently the access relation must be an equivalence).

Because a deductive proof does not generate new information, the
conclusion must be ``hidden'' in the premises \cite[Section
1.2(5)]{Eder&Ramharter15} and here it is hidden in the combination of
Anselm's Principle and the modal axioms.  But all of these are well
motivated, so the argument is quite satisfying.  In contrast (2),
particularly when written $\Diamond g \supset \Box g$, is not
persuasive as a premise and it reduces the argument to triviality.
Similarly (4), particularly when written $\Diamond g \supset g$, is
unpersuasive and also trivializes the argument.  

Bi{\l}at himself eventually rejects (4), but defends (2) and claims
that the standard version (i.e., (3) with Modal Axioms B or 5 and T)
is ``philosophically inferior'' to (2) with Modal Axiom T.

Before concluding, we should note that Bi{\l}at also mentions a
premise, which he attributes to \cite{Oppy96}, that we will call
(5): $\Diamond\Box g$.  This premise is able to discharge the
conclusion $g$ using Modal Axiom B but without using (1).  However, it
is entailed by the conjunction of Premises (1) and (3), so it can be
considered a lemma, having those two premises ``built in.''  Unlike
the other alternative premises, this one does not trivialize the
argument (since it still requires Modal Axiom B), and seems no less
plausible as a premise than (1).

We conclude that Bi{\l}at's alternative premises that simplify the
Modal Ontological Argument by using fewer modal axioms than those that
employ a traditional formulation of Anselm's Principle, such as (3),
are the result of applying the discarded modal axioms to (3) and
therefore covertly build in those axioms (i.e., they are not
independent premises but lemmas, proved using the discarded modal
axioms).  Furthermore, the alternative premises are unpersuasive as
premises, and they trivialize the argument.


\renewcommand{\refname}{Additional References}

\end{document}